\documentstyle[prl,aps]{revtex}
\newcommand{\be}{\begin{equation}}
\newcommand{\ee}{\end{equation}}
\newcommand{\ba}{\begin{eqnarray}}
\newcommand{\ea}{\end{eqnarray}}
\begin{document}
\draft
\twocolumn
\title{An algebraic proof on the finiteness of Yang-Mills-Chern-Simons 
theory in $D=3$}
\author{O.M. Del Cima$^a$\thanks{
Supported by the Fonds zur F\"orderung der Wissen\-schaftlichen Forschung
(FWF) under the contract number P11654-PHY.}, 
D.H.T. Franco$^{b}$\thanks{
Supported by the Conselho Nacional de Desenvolvimento Cient{\'\i}fico e
Tecnol{\'o}gico (CNPq).}, 
J.A. Helay\"el-Neto$^{b\dagger}$ and 
O. Piguet$^{c\dagger}$\thanks{Supported in part by the Swiss National 
Science Foundation.}\thanks{
On leave of absence from D\'epartement de Physique Th\'eorique,
Universit\'e de Gen\`eve, 24 quai E. Ansermet - CH-1211 - Gen\`eve 4 -
Switzerland.}}
\address{{\normalsize {\it $^a$Institut f\"ur Theoretische Physik,}}\\
Technische Universit\"at Wien (TU-Wien),\\
Wiedner Hauptstra{\ss}e 8-10 - A-1040 - Vienna - Austria.\\
{\normalsize {\it $^b$Centro Brasileiro de Pesquisas F\'\i sicas (CBPF),}}\\
Departamento de Teoria de Campos e Part\'\i culas (DCP), \\
Rua Dr. Xavier Sigaud 150 - 22290-180 - Rio de Janeiro - RJ - Brazil.\\
{\normalsize {\it $^c$Universidade Federal do Esp\'\i rito Santo (UFES),}}\\
CCE, Departamento de F\'\i sica, \\
Campus Universit\'ario de Goiabeiras - 29060-900 - Vit\'oria - ES - Brazil.}
\date{\today}
\maketitle
\begin{abstract}
A rigorous algebraic proof of the full finiteness in all orders of
perturbation theory is given for the Yang-Mills-Chern-Simons theory in a
general three-dimensional Riemannian manifold. We show the validity of a
trace identity, playing the role of a local form of the Callan-Symanzik
equation, in all loop orders, which yields the vanishing of the $\beta$
-functions associated to the topological mass and gauge coupling constant as
well as the anomalous dimensions of the fields.
\end{abstract}

\bigskip

The finiteness of the Yang-Mills-Chern-Simons (YM\-CS) theory~\cite
{deser1,pisarski,deser2,martin,giavarini} in $D=3$ has been pursued since
its evidence was first detected by one-loop order calculations~\cite
{pisarski,deser2}, and later on up to two-loops~\cite{giavarini}. 
Recently, the finiteness of the $N=1$ super-YMCS 
theory~\cite{ruiz} has been shown. A partial proof on the finiteness
of $N=2$ super-YMCS theory in the Wess-Zumino gauge is given 
in~\cite{maggiore}. Since the pure Chern-Simons (CS) theory 
is finite to all orders in perturbation theory~\cite{blasi}, 
two recent papers~\cite{lemes1,lemes2} have claimed the 
equivalence of the YMCS theory with a pure CS theory 
at the quantum level to argued the finiteness of 
YMCS up to field amplitude renormalizations. Very recently a proof has been 
done of vanishing $\beta$-functions associated to non-invariant BRS local 
terms~\cite{barnich}.

In this letter we present a rigorous proof of the full finiteness of the
YMCS theory in a general three-dimensional Riemannian manifold. 
The approach we propose here to quantum 
scale invariance of the YMCS is based on the 
energy-momentum (EM) tensor trace identity, playing the 
role of a local form of the Callan-Symanzik equation. It means exact 
quantum scale invariance, with vanishing $\beta$-functions and 
anomalous dimensions as well. 

The same technique~\cite{cfhp1} has been used to prove the full 
finiteness of the BF-Yang-Mills theory in $D=3$~\cite{cfhp2}. 
To give such a proof on the full 
quantum scale invariance of YMCS we use the algebraic renormalization 
method~\cite{becchi,piguet1,piguet2}. It is based on the 
BRS-formalism~\cite{becchi} together with the Quantum Action 
Principle~\cite{lowenstein}, which leads to a regularization
independent scheme. We think indeed that, due to the presence of the
antisymmetric Levi-Civita tensor, it is difficult to establish an invariant
regularization scheme without encountering problems at some or other stage
of the argument.

Since we are working with an external curved dreibein, 
our results hold for a curved manifold, as long as its topology
remains that of flat ${\cal R}^3$. This allows us to use the 
general results of renormalization 
theory~\cite{lowenstein,zimmermann} established in flat space. 

The three-dimensional space-time is a Riemannian manifold described by 
a dreibein field $e_\mu^m$. The spin connection $\omega_\mu^{mn}$ 
depends on the dreibein due to the vanishing
torsion condition. The metric tensor reads 
$g_{\mu\nu}=\eta_{mn}e_\mu^m e_\nu^n$, with $\eta_{mn}$ being the
tangent flat space metric. We denote by $e$ the determinant of $e_\mu^m$.

The YMCS classical action (in the Landau gauge) in a three-dimensional 
curved manifold reads:
\begin{eqnarray}
\Sigma&=&\int d^3x~\left\{-\frac e4 F_{\mu\nu}^aF^{a\mu\nu} +
m\varepsilon^{\mu\nu\rho}(A_\mu^a\partial_\nu A_\rho^a \right. + \nonumber \\
&&+~\frac g3f_{abc}A_\mu^aA_\nu^bA_\rho^c) - e g^{\mu\nu}
(\partial_\mu b_aA_\nu^a+\partial_\mu \bar{c}_aD_\nu c^a) +  \nonumber \\
&&+\left. (A_a^{*\mu}sA_\mu^a+c_a^{*}sc^a)\right\}~~,
\label{g-inv-action}
\end{eqnarray}
where $m$ is the topological mass~\cite{deser1} and $g$ is the gauge
coupling constant. The field strength is defined
as $F_{\mu\nu}^a=\partial_\mu A_\nu^a-\partial_\nu A_\mu^a+
gf_{abc}A_\mu^bA_\nu^c$ and $c^a$, $\bar{c}^a$ and $b^a$ are
the ghost, the antighost and the Lagrange multiplier fields, respectively. 
$A_a^{*\mu}$ and $c_a^{*}$ are the ``antifields'' (tensorial densities) 
coupled to the nonlinear
variations of the fields $A_\mu^a$ and $c^a$ under BRS transformations $s$: 
\begin{eqnarray}
sA_\mu^a&=&-D_\mu c^a\equiv-(\partial_\mu c^a+gf_{abc}A_\mu^bc^c)~~,  
\nonumber \\
sc^a &=&\frac g2f_{abc}c^bc^c~,~~s{\bar{c}}^a=b^a~,~~sb^a=0~~.
\label{BRSnonab1}
\end{eqnarray}

The action (\ref{g-inv-action}) is also invariant under diffeomorphisms 
\begin{equation}
\delta_{{\rm diff}}^{(\varepsilon)}\Phi={\cal L}_\varepsilon \Phi~~,
\label{diff-fields}
\end{equation}
where $\Phi=(A_\mu^a,e_\mu^m,c^a,b^a,{\bar{c}}^a,A_a^{*\mu},c_a^{*})$ and 
${\cal L}_\varepsilon$ the Lie derivative along the
infinitesimal vector field $\varepsilon^\mu$; and under infinitesimal 
local Lorentz transformations
\begin{equation}
\delta_{{\rm Lorentz}}^{(\lambda)}\Phi=\frac 12\lambda_{mn}\Omega^{mn}\Phi~~,
~~~\Phi=\mbox{any field}~~,  \label{lor-fields}
\end{equation}
with $\Omega^{[mn]}$ acting on $\Phi$ as a Lorentz matrix in the appropriate 
representation.

The BRS invariance of the action is expressed in a functional way by the
Slavnov-Taylor (ST) identity 
\begin{equation}
{\cal S}(\Sigma)={\int}d^3x\left({\frac{\delta\Sigma}{\delta A_a^{*\mu}}}
{\frac{\delta\Sigma}{\delta A_\mu^a}}+{\frac{\delta\Sigma}{\delta c_a^{*}}}
{\frac{\delta\Sigma}{\delta c^a}}+b^a{\frac{\delta\Sigma}
{\delta\bar{c}^a}}\right)=0 \label{slavnonab1}
\end{equation}
where the corresponding linearized ST operator reads 
\begin{eqnarray}
{\cal B}_\Sigma &=&{\int}d^3x\left({\frac{\delta\Sigma}{\delta A_a^{*\mu}}}
{\frac \delta{\delta A_\mu^a}}+{\frac{\delta\Sigma}{\delta A_\mu^a}}
{\frac \delta{\delta A_a^{*\mu}}}+{\frac{\delta\Sigma}{\delta c_a^{*}}}
{\frac \delta{\delta c^a}}\right. + \nonumber \\
&&+\left. {\frac{\delta \Sigma}{\delta c^a}}{\frac \delta{\delta c_a^{*}}}+
b^a{\frac \delta{\delta \bar{c}^a}}\right)~~.  \label{linear}
\end{eqnarray}

The operators ${\cal S}$ and ${\cal B}$ obey the following nilpotency 
identities:  ${\cal B}_{{\cal F}}~{\cal S}({\cal F})=0$ $\forall{\cal F}$, 
and $({\cal B}_{{\cal F}})^2=0$ if $~{\cal S}({\cal F})=0$. In particular,
since the action $\Sigma $ obeys the ST identity 
(\ref{slavnonab1}), we have the nilpotency property $({\cal B}_\Sigma)^2=0$.

In addition to the ST identity (\ref{slavnonab1}), the action 
(\ref{g-inv-action}) satisfies the constraints: the Landau gauge 
condition 
\begin{equation}
{\frac{\delta\Sigma}{\delta b_a}}=\partial_\mu (eg^{\mu\nu}A_\nu^a)~~;  
\label{landau1}
\end{equation}
and the ``antighost equation'' (in the Landau gauge~\cite{blasi}) 
\begin{equation}
{\bar{{\cal G}}}^a\Sigma={\int}d^3x\left({\frac \delta {\delta c^a}}+
gf^{abc}\bar{c}_b{\frac \delta {\delta b^c}}\right)\Sigma=
\Delta_{{\rm cl}}^a~~;  \label{antighost}
\end{equation}
with $\Delta_{{\rm cl}}^a=g{\int}d^3xf^{abc}(A_b^{*\mu}A_{c\mu}-c_b^{*}c_c)$. 
Note that the right-hand side of (\ref{antighost}) being linear in the
quantum fields, will not be submitted to renormalization.

The Ward (W) identities for the diffeomorphisms (\ref{diff-fields}) and 
the local Lorentz transformations (\ref{lor-fields})
read: 
\begin{equation}
{\cal W}_{X}\Sigma={\int}d^3x\sum_{{\rm all~fields}}^{}
\delta_{X}\Phi 
{\frac{\delta \Sigma}{\delta \Phi}}=0~~, \label{wdiffeolorentz}
\end{equation}
where $X=({\rm diff,Lorentz})$.

Commuting (\ref{slavnonab1}) and (\ref{landau1}) we obtain  
\begin{equation}
{\cal G}^a\Sigma={{\frac{\delta \Sigma}{\delta \bar{c}_a}}}+\partial_\mu
\left(eg^{\mu\nu}{{\frac{\delta \Sigma}{\delta A_a^{*\nu}}}}\right)=0~~,  
\label{ghost1}
\end{equation}
which is the ``ghost equation''~\cite{piguet2}. It implies that 
the theory depends on the field $\bar{c}_a$ and on the
antifield $A_a^{*\mu}$ through the combination ${\hat{A}}_a^{*\mu}=A_a^{*\mu}+
eg^{\mu\nu}\partial_\nu\bar{c}_a$.

Moreover, the action (\ref{g-inv-action}) is invariant under the rigid gauge
transformations, given by the W identity 
\begin{equation}
{\cal W}_{{\rm rigid}}^a\Sigma={\int}d^3x
{\sum_{\phi=A,c,\bar{c},b,A^{*},c^{*}}^{}}f^{abc}
\phi_b{{\frac{\delta \Sigma}{\delta \phi ^c}}}=0~~,
\label{crigidcondnonab}
\end{equation}
by anticommuting (\ref{slavnonab1}) and (\ref{antighost}).

In order to give a proof of the renormalizability of (\ref{g-inv-action}),
we have to show that all constraints defining the classical theory also hold 
at the quantum level, {\it i.e.} that we can construct a renormalized vertex 
functional $\Gamma=\Sigma +{\cal O}(\hbar)$, obeying the same constraints 
and coinciding with the classical action at order zero in $\hbar$.

The first point to be checked is the power-counting renormalizability. 
The ultraviolet dimension, as well as the ghost
number and the Weyl dimension of all fields and antifields are collected in
Table I.

In order to explicitly find the possible renormalizations and anomalies of
the theory, we can use the following result~\cite{maggiore}: the degree of
divergence of a 1-particle irreducible Feynman graph $\gamma$ is given by 
\begin{equation}
d(\gamma)=3-\sum\limits_{\tilde{\Phi}=\Phi,g}d_{\tilde{\Phi}} 
N_{\tilde{\Phi}}~~~,~~~~~\mbox{with}~~~~~d_g=\frac 12~~.
\end{equation}
Here $N_\Phi$ is the number of external lines of $\gamma$ corresponding to
the field $\Phi$, $d_\Phi$ is the dimension of $\Phi$ as given in Table I,
and $N_g$ is the power of the coupling constant $g$ in the integral
corresponding to the diagram $\gamma$. In order to 
apply the known results on the quantum action principle~\cite
{lowenstein} to the present situation, we have considered $g$ as an external
field of dimension $\frac 12$. 

Thus, including the dimension of $g$ into the calculation, we may state that
the dimension of the counterterms of the action is bounded by 3. However,
since they are generated by loop graphs, they are of order 2 in $g$ at
least. This means that, not taking now into account the dimension of $g$, we
can conclude that their real dimension is bounded by 2. The same holds for
the possible breakings of the ST identity.

The second point to be discussed concerns about the functional identities to
be obeyed by the vertex functional $\Gamma$. The
gauge condition (\ref{landau1}), antighost
equation (\ref{antighost}), ghost equation (\ref{ghost1}) 
as well as rigid gauge invariance 
(\ref{crigidcondnonab}) can be easily shown to hold at all orders, {\it i.e.}
are not anomalous~\cite{piguet2}. The validity to all orders of the W
identities of diffeomorphisms and local Lorentz will be assumed in the
following: the absence of anomalies for them has been proved 
in~\cite{brandt,barnich1} for the class of manifolds we are considering here.

It remains now to show the possibility of implementing the ST
identity (\ref{slavnonab1}) for the vertex functional $\Gamma$. As it is
well known~\cite{piguet2}, this amounts to study the cohomology of the
nilpotent operator ${{\cal B}_\Sigma}$, defined by (\ref{linear}), in the
space of local integrated functionals $\Delta$ of the fields
involved in the theory. The cohomology classes of ${{\cal B}_\Sigma}$ are
defined such that $\Delta$ and $\Delta +{{\cal B}_\Sigma }\hat{\Delta}$
belong to the same equivalence class. The set of these classes is called the
cohomology group ${\cal H}^p({{\cal B}_\Sigma})=
{\cal Z}^p({{\cal B}_\Sigma})/{\cal Q}^p({{\cal B}_\Sigma})$; 
${\cal Z}^p({{\cal B}_\Sigma})$ being the space of cocycles (the
nontrivial part of the general solution) and ${\cal Q}^p({{\cal B}_\Sigma})$ 
being the space of coboundaries (BRS-variation) both of
ghost number $p$. The cohomological group ${\cal H}^0({{\cal B}_\Sigma})$ 
constitutes the non-trivial invariants of the theory, {\it i.e.} the
arbitrary invariant counterterms we can add to the action at each order of
perturbation theory which correspond to the renormalization of the physical
parameters (coupling constants and masses), whereas 
${\cal Q}^0({{\cal B}_\Sigma})$ represents the non-physical renormalizations 
(field amplitudes). On the other hand, ${\cal H}^1({{\cal B}_\Sigma})$
is related to the possible anomalies.

In the both cases, ${\cal H}^0$ and ${\cal H}^1$, the
super-renormalizability by power-counting restricts the dimension of the
integrand of $\Delta $ to 2. Moreover, the constraints 
(\ref{landau1}--\ref{crigidcondnonab}), 
valid now for the vertex functional $\Gamma$, imply for $\Delta$ the 
conditions 
\begin{equation}
{\frac \delta {\delta b_a}}\Delta=
{\int}d^3x{\frac \delta {\delta c^a}}\Delta=
{\cal G}^a\Delta=
{\cal W}_X\Delta=0~~,\label{cond-delta}
\end{equation}
where $X=({\rm diff,Lorentz,rigid})$.

It has been proven in quite generality~\cite{barnich1,barnich2} that in such
a gauge theory the cohomology in the sector of ghost number 1 is independent
of the external fields (antifields). We can thus restrict the field 
dependence of $\Delta$ to $A_\mu^a$ and $c^a$, with the dependence on $c^a$ 
being through its derivatives due to the second of the constraints 
(\ref{cond-delta}).

Beginning with the anomalies, we know~\cite{brandt,barnich2} that, 
in three dimensions, the cohomology in this sector
is empty, up to possible terms in the Abelian ghosts. However, they can be
seen, by using the arguments of~\cite{bandelloni}, not to contribute to the
anomaly, due to their freedom or soft coupling. We thus
conclude to the absence of gauge anomaly, hence to the validity of the
ST identity (\ref{slavnonab1}) to all orders for the vertex
functional $\Gamma$.

Going now to the sector of ghost number 0, {\it i.e.} looking for the
arbitrary invariant counterterms which can be freely added to the action at
each order. According to the above discussion the counterterm is at least of
order $g^2$. Thus, the most general expression for the nontrivial part of 
$\Delta$ reads 
\begin{equation}
\Delta_{{\rm phys.}}=z_mm\frac \partial {\partial m}\Sigma ~~,
\label{cterm0}
\end{equation}
where $z_m$ is an arbitrary parameter. Eq.(\ref{cterm0}) 
shows that, {\it a priori}, 
only the parameter $m$ can get radiative corrections. 
This means that the $\beta_g$-function related to 
the gauge coupling constant $g$ is vanishing to all orders of perturbation 
theory, and the anomalous dimensions of the fields as well. 
This concludes the proof of the
renormalizability of the theory: all functional identities hold without
anomaly and the renormalizations might only affect the CS coupling, 
{\it i.e.} the topological mass $m$. But the latter turns out to
be not renormalized, too. We shall indeed show that its corresponding 
$\beta_m$-function vanishes at all orders, which yields the full finiteness 
of the YMCS theory in a three-dimensional Riemmanian manifold.

Now, a precise study on the quantum scaling properties of the YMCS theory 
demands a local version of the Callan-Symanzik equation. Its local form 
arises from the ``trace identity''. It will be useful to exploit the fact 
that the integrand of the CS action is not gauge invariant, in 
spite of its integral be. This strong constraint upon the quantum insertions, 
together with the others, will guarantee that no insertions survive at all,
therefore, as a consequence, the vanishing of the topological mass 
$\beta_m$-function. Above all, let us introduce the EM tensor, defined 
as the following tensorial quantum insertion: 
\begin{equation}
{\Theta}_\nu^{~\mu}\cdot \Gamma=e^{-1}e_\nu^{~m}~\frac{\delta \Gamma}
{\delta e_\mu^{~m}}~~.  \label{theta}
\end{equation}

The integral of the trace of the tensor $\Theta_\nu^{~\mu}$ 
\begin{equation}
{\int}d^3x~e~\Theta_\mu^{~\mu}={\int}d^3x~e_\mu^{~m}\frac{\delta \Sigma}
{\delta e_\mu^{~m}}\equiv {\cal N}_e\Sigma  \label{int-theta}
\end{equation}
follows from the identity 
\begin{equation}
{\cal N}_e\Sigma =\left[\sum\limits_{{\rm all~fields}}d_W(\Phi)
{\cal N}_\Phi + m\partial_m+\frac 12g\partial_g\right] \Sigma~~,  \label{Ne}
\end{equation}
where the operators ${\cal N}_\Phi={\int}d^3x~\Phi{\frac \delta {\delta \Phi}}$
are the counting operators and $d_W(\Phi)$ the Weyl dimension
(see Table I) of the field $\Phi$. It should be noticed that 
(\ref{int-theta}) is nothing else than the W identity for the rigid Weyl 
symmetry~\cite{iorio}.

The trace $\Theta_\mu^{~\mu}\cdot\Gamma$ turns out to be vanishing
up to total derivatives and dimensionful couplings, in the
classical approximation, due to the field equations, which means that 
(\ref{theta}) is the improved EM tensor. It is easy to check that
from the classical action the following equation holds 
\begin{equation}
w\Sigma \equiv \left[ e_\mu^{~m}\frac \delta {\delta e_\mu^{~m}}-
\sum\limits_{{\rm all~fields}}d_W(\Phi) \Phi {\frac \delta {\delta \Phi}}
\right] \Sigma=\Lambda~~,  \label{ym-cl-tr1}
\end{equation}
with $\Lambda$ being ${\cal B}_\Sigma$-invariant. 
It should be pointed out that $\Lambda$ is the effect of the 
breaking scale invariance caused by the dimensionful couplings. 
In fact, it is a soft breaking, since its dimension is lower than 3 
(the dimensions of $m$ and $g$ are not taken into account)

To promote the trace identity (\ref{ym-cl-tr1}) to the quantum level, we
first note that the following conditions for the insertion $w\Gamma$ hold 
\begin{eqnarray}
{\cal B}_\Gamma w(x)\Gamma&=&0~,
~~{\bar{{\cal G}}}^aw(x)\Gamma=
\frac 12 \frac{\delta \Gamma}{\delta c_a(x)}~~, \nonumber \\
\frac \delta {\delta b_a(y)}w(x)\Gamma&=&-\frac 32\partial_\mu^x
\delta(x-y)(eg^{\mu\nu}A_\nu^a)(y)~~,  \label{cba'} \\
{\cal G}^a(y)w(x)\Gamma&=&\frac 32\partial_\mu^x\delta(x-y)
\left(eg^{\mu\nu}\frac{\delta \Gamma}{\delta A_a^{*\nu}}\right) (y)~~,
\nonumber
\end{eqnarray}
where we use again the fact that the constraints (\ref{landau1}), 
(\ref{antighost}) and (\ref{ghost1}) can be maintained at the quantum 
level~\cite{piguet2}.

The quantum version of (\ref{ym-cl-tr1}) is written as 
\begin{equation}
w\Gamma=\Lambda\cdot\Gamma+\Delta\cdot\Gamma ~~,
\label{ym-quant-trace}
\end{equation}
where $\Lambda\cdot\Gamma$ is some quantum extension of the classical
insertion $\Lambda$, subjected to the same constraints (\ref{cba'}) 
as $w\Gamma$ (see~\cite{cfhp2}). 
It follows that the insertion $\Delta\cdot\Gamma$ defined by 
(\ref{ym-quant-trace}) obeys the homogeneous constraints 
\begin{equation}
{\cal B}_\Gamma [\Delta\cdot\Gamma]=
\frac \delta {\delta b_a}[\Delta\cdot\Gamma]=
{\bar{{\cal G}}}^a[\Delta\cdot\Gamma]=
{\cal G}^a[\Delta\cdot\Gamma]=0 \label{cond1}
\end{equation}
beyond the conditions of invariance or covariance under 
${\cal W}_{{\rm diff}}$, ${\cal W}_{{\rm Lorentz}}$ and 
${\cal W}_{{\rm rigid}}$.

By power-counting the insertion $\Delta\cdot\Gamma$ has dimension 3, but
being an effect of the radiative corrections, it possesses a factor $g^2$ at
least, and thus its effective dimension is at most 2. It turns out that
there is no insertion obeying all these constraints, the power-counting
selects the CS Lagrangian, but the latter is not BRS
invariant. Therefore, $\Delta\cdot\Gamma=0$: there is no radiative
correction to the insertion $\Lambda\cdot\Gamma$ describing the breaking
of scale invariance. It follows that (\ref{ym-quant-trace}) becomes 
\begin{equation}
e~{\Theta}_\mu^{~\mu}\cdot\Gamma=
\sum\limits_{{\rm all~fields}}d_W(\Phi)\Phi
{\frac {\delta\Gamma} {\delta \Phi}}+\Lambda\cdot\Gamma~~.  
\label{trace-id-tilde}
\end{equation}
This local trace identity leads to a Callan-Symanzik equation (see Section 6
of~\cite{cfhp1}): 
\begin{equation}
\left(m\partial_m+\frac 12g\partial_g\right)\Gamma=
\int d^3x~\Lambda\cdot\Gamma~~,
\end{equation}
where no radiative effect contributes, that results in the vanishing 
$\beta$-functions associated to the parameters $g$ ($\beta_g$) and $m$ 
($\beta_m$) as well as the anomalous dimensions of the fields. The scale 
invariance remains affected only by the soft breaking $\Lambda$. We have 
thus shown that there is no renormalization at all: the 
Yang-Mills-Chern-Simons theory in $D=3$ is UV finite.

In conclusion, the method we have presented here has been allowed us to give
a rigorous proof based on general theorems of renormalization theory on the
full finiteness of the YMCS theory in a three-dimensional Riemannian
manifold at all orders in perturbation theory. Also, this method turns out
possible the identification of the real causes that are from behind the
finiteness of the YMCS theory.

\begin{table}[tbp]
\par
\begin{center}
\begin{tabular}{|c||c|c|c|c|c|c|c|}
\hline
& $A_\mu$ & $b$ & $c$ & ${\overline{c}}$ & $A^{*\mu}$ & $c^{*}$ & $g$ \\ 
\hline\hline
$d$ & $1/2$ & $3/2$ & $-1/2$ & $3/2$ & $5/2$ & $7/2$ & $1/2$ \\ \hline
$\Phi\Pi$ & $0$ & $0$ & $1$ & $-1$ & $-1$ & $-2$ & $0$ \\ \hline
$d_W$ & $-1/2$ & $3/2$ & $-1/2$ & $3/2$ & $1/2$ & $1/2$ & $1/2$ \\ \hline
\end{tabular}
\end{center}
\label{table}
\caption{Ultraviolet dimension $d$, ghost number $\Phi \Pi $ and Weyl
dimension $d_W$.}
\end{table}

\underline{Acknowledgements}: The authors thank Prof. Manfred Schweda and 
Dr. Emmanuel A. Pereira for the critical reading of the manuscript. One of 
the authors (O.M.D.C.) dedicates this work to his wife, Zilda Cristina, to 
his daughter, Vittoria, and to his son, Enzo, who was born on March 27th, 
1998. 

\noindent
* E-mail: delcima@tph73.tuwien.ac.at . \\
\dag~E-mail: dfranco@cbpfsu1.cat.cbpf.br . \\
\ddag~E-mail: helayel@cbpfsu1.cat.cbpf.br . \\
\S~E-mail: piguet@cce.ufes.br .

\end{document}